\documentclass[12pt]{iopart}

\usepackage{graphicx}
\usepackage{amssymb}
\usepackage{iopams}  

\def\beq{\begin{equation}}
\def\eeq{\end{equation}} 
\def\br{\begin{eqnarray}}
\def\er{\end{eqnarray}}
\def\benu{\begin{enumerate}}
\def\eenu{\end{enumerate}}
\def\nn{\nonumber} 
\def\pa{{\partial}}
\def\l{\left}
\def\r{\right}

\begin{document}


\title{On the evolution of tachyonic perturbations at 
super-Hubble scales}
\author{Rajeev Kumar Jain}
\address{Harish-Chandra Research Institute, Chhatnag Road,
Jhunsi, Allahabad~211~019, India.}
\ead{rajeev@mri.ernet.in}
\author{Pravabati Chingangbam}
\address{Korea Institute for Advanced Study,
207--43 Cheongnyangni 2-dong, Dongdaemun-gu, Seoul 130-722, Korea.}
\ead{prava@kias.re.kr}
\author{L.~Sriramkumar}
\address{Harish-Chandra Research Institute, Chhatnag Road,
Jhunsi, Allahabad~211~019, India.}
\ead{sriram@mri.ernet.in}
\begin{abstract}
In the slow-roll inflationary scenario, the amplitude of the curvature 
perturbations approaches a constant value soon after the modes leave
the Hubble radius.
However, relatively recently, it was shown that the amplitude of the
curvature perturbations induced by the canonical scalar field can 
{\it grow}\/ at super-Hubble scales if there is either a transition 
to fast roll inflation or if inflation is interrupted for some period 
of time.
In this work, we extend the earlier analysis to the case of a 
non-canonical scalar field described by the Dirac-Born-Infeld 
action.
With the help of a specific example, we show that the amplitude of 
the tachyonic perturbations can be {\it enhanced or  suppressed}\/ 
at super-Hubble scales if there is a transition from slow roll to 
fast roll inflation.   
We also illustrate as to how the growth of the entropy perturbations 
during the fast roll regime proves to be responsible for the change in
the amplitude of the curvature perturbations at super-Hubble scales.
Furthermore, following the earlier analysis for the canonical scalar 
field, we show that the power spectrum evaluated in the long wavelength 
approximation matches the exact power spectrum obtained numerically 
very well. 
Finally, we briefly comment on an application of this phenomenon.
\end{abstract}
\maketitle


\section{Introduction}

It is well-known that, in the inflationary scenario, the 
amplitude of the curvature perturbations approaches a 
constant value soon after the modes leave the Hubble radius 
(see, for example, either of the following texts~\cite{texts} 
or one of the following reviews~\cite{reviews}).  
But, what is not so commonly known is that this result is true 
{\it only}\/ in slow roll or power law inflation and, in fact, 
the curvature perturbations can be amplified at super-Hubble 
scales if there is either a period of fast roll inflation or if 
there is a break in inflation. 
For the case of curvature perturbations induced by the canonical 
scalar field, this behavior was first noticed a few years 
back~\cite{leach01a} and a general criterion for such an 
amplification to occur was also obtained~\cite{leach01b}. 
Our aim in this work is to extend the earlier 
analysis~\cite{leach01a,leach01b} to the case of curvature 
perturbations generated by a non-canonical scalar field. 
Interestingly, we find that, in addition to enhancing the amplitude 
of the curvature perturbations of a certain range of modes at 
super-Hubble scales, a period of deviation from slow roll inflation 
also leads to the suppression of the amplitude of another range of 
modes.

The non-canonical scalar field that we shall consider is the 
one that is described by the Dirac-Born-Infeld (DBI, hereafter) 
action. 
Such scalar fields generically arise in the study of the dynamics 
of D-branes in string theory~\cite{dbi1}.
In particular, the tachyon, which refers to an unstable DBI scalar
field that rolls down from the maxima of its potential near the origin 
to its minima at infinity, captures the essential dynamical features 
of the decay process of unstable branes (for the original discussion, 
see Refs.~\cite{sen02}). 
Various cosmological applications of the tachyon have been studied 
extensively in the literature (see, for instance, Refs.~\cite{dbi2} 
and references therein) and, in particular, it has proved to be a 
fertile ground for inflationary model building (see, for example, 
Ref.~\cite{dbicosmology} and references therein). 
It is also possible to have DBI scalar fields whose potential has 
a global minimum at the origin so that the field rolls down from 
a large value towards the origin (see, for example, 
Ref.~\cite{garousi03}; for cosmological applications of such fields, 
see, for instance, Refs.~\cite{garousi04}).
We shall consider such a scenario in this work. 
It should be pointed out here that the DBI scalar field falls under 
a broader class of non-canonical scalar field models that are often 
referred to as the k-inflation models~\cite{kinflation}. 

With the help of a specific example, we shall illustrate that the 
amplitude of the curvature perturbations induced by the DBI scalar 
field can be enhanced or suppressed (when compared to its value 
at Hubble exit) at super-Hubble scales if there is a period of 
deviation from slow roll inflation.
We shall also show that it is the growth of entropy perturbations 
during such a transition that turns out to be responsible for the 
change in the amplitude of the tachyonic perturbations\footnote{Actually, 
as we mentioned above, the term `tachyon' refers to a DBI scalar 
field that is described by a potential with a maxima near the origin.
The specific potential we shall consider in this paper does not have 
this feature. 
Nevertheless, we shall often use the terms `tachyon' and `tachyonic
perturbations' as convenient shorthands for referring to the 
DBI scalar field and to the perturbations induced by it, 
respectively.}. 
Moreover, following the earlier analysis for the canonical scalar 
field, we shall show that the power spectrum evaluated in the long 
wavelength approximation and the exact power spectrum obtained 
numerically match rather well.
We shall also very briefly comment on an interesting application 
of this phenomenon.

The remainder of this paper is organized as follows.
In the following section, we shall briefly summarize the essential 
background equations and the equations describing the scalar 
perturbations generated by the DBI scalar field.
In Section~\ref{sec:ashsse}, we shall illustrate the amplification
or suppression of the curvature perturbations at super-Hubble scales 
when there is a transition from slow roll to fast roll and back with 
the help of a specific example.
We shall also show as to how the entropy perturbations grow during 
the period of fast roll inflation which in turn act as the source 
for the evolution of the curvature perturbations at super-Hubble
scales.
In Section~\ref{sec:cep}, we extend an earlier result for the 
canonical scalar field to the tachyonic case and show that the
power spectrum calculated in the long wavelength approximation 
agrees quite well with the exact power spectrum evaluated 
numerically.
Finally, in Section~\ref{sec:summary}, we close by commenting on an
important application of our result.

Before we proceed further, a few words on the conventions and notations 
we shall adopt are in order. 
We shall set $\hbar=c=1$ and work with the metric signature of 
$(+,-,-,-)$.
We shall express the various quantities in terms of either the cosmic 
time~$t$ or the conformal time $\eta$, as is convenient.
An overdot shall denote differentiation with respect the cosmic time and 
an overprime shall denote differentiation with respect to the conformal 
time.
It is useful to note here that, for any given function, say, $f$, 
${\dot f}=(f'/a)$ and ${\ddot f}=\l[\l(f''/a^2\r)-\l(f'\, 
a'/a^{3}\r)\r]$, where $a$ is the scale factor describing the 
Friedmann background.


\section{Essential results from cosmological perturbation 
theory}\label{sec:cpt}

In this section, we shall briefly outline essential cosmological
perturbation theory.
We shall quickly summarize the equations describing the Friedmann 
background and the tachyonic perturbations that we will require for 
our discussion.
 
\subsection{The background equations}

Consider a $(3 + 1)$-dimensional, spatially flat, smooth, Friedmann 
universe described by line element
\beq
ds^2 = dt^2-a^{2}(t)\, d{\bf x}^2 
= a^{2}(\eta)\, \l(d\eta^{2} - d{\bf x}^2\r),\label{eq:frwle}
\eeq
where $t$ is the cosmic time, $a(t)$ is the scale factor and $\eta=\int 
dt/a(t)$ denotes the conformal time.  
If $\rho$ and $p$ denote the energy density and the
pressure of the smooth component of the matter field that is 
driving the expansion, then the Einstein's equations for the 
above line-element lead to the following Friedmann equations 
for the scale factor~$a(t)$:
\beq
\l(\frac{\dot a}{a}\r)^{2}
= \l(\frac{8\pi\, G}{3}\r)\, {\rho}\qquad{\rm and}\qquad
\l(\frac{\ddot a}{a}\r)
= -\l(\frac{4\pi\, G}{3}\r)\, 
\l({\rho}+3\, p\r).
\eeq
For the case of a DBI scalar field, say, $T$, described by the 
potential~$V(T)$, the background energy density $\rho$ and pressure 
$p$ are given by 
\beq
\rho = \l(\frac{V(T)}{\sqrt{1 - {\dot T}^2}}\r)
\qquad{\rm and}\qquad
p = -\l(V(T)\, \sqrt{1 - {\dot T}^2}\r).
\eeq
Also, in the spatially flat, Friedmann background that we shall 
be considering, the equation of motion for the tachyon~$T$ is 
given by
\beq
\l(\frac{\ddot T}{1-\dot{T}^2}\r)
+(3\, H)\;\dot{T}+\l(\frac{V_{T}}{V}\r)=0, \label{eq:emT}
\eeq
where $H=({\dot a}/a)$ denotes the Hubble parameter and $V_{T}
\equiv \l(dV/dT\r)$.


\subsection{The scalar perturbations}

\subsubsection{Generic results}

If we now take into account the scalar perturbations to the background 
metric~(\ref{eq:frwle}), then the Friedmann line-element, in general,  
can be written as~\cite{texts,reviews,sasaki}
\beq
\fl\quad\;
{\rm d}s^2 
= a^2(\eta)\, \l[(1+2 A)\,{\rm d}\eta ^2 - 2\, (\pa_{i} B)\,{\rm d}x^i\, 
{\rm d}\eta - \l((1-2 \psi)\, \delta _{ij}+ 2\, \pa_{i}\pa_{j}E \r)\, 
{\rm d}x^i\, {\rm d}x^j\r],\label{eq:frwlesp}
\eeq
where $A$, $B$, $\psi$ and $E$ are the scalar functions that describe 
the perturbations.
The two gauge-invariant Bardeen variables that characterize the 
two degrees of freedom describing the scalar perturbations are 
given by~\cite{texts,reviews,sasaki}
\beq
\Phi 
\equiv A +\l(\frac{1}{a}\r)\, \l[(B-E^{\prime})\, a\r]^{\prime}
\qquad {\rm and}\qquad 
\Psi 
\equiv \psi - {\cal H}\, \l(B-E^{\prime}\r).\label{eq:PhiPsi}
\eeq
In the absence of anisotropic stresses, as it is in the case of the
scalar field sources that we are interested in, it can be readily 
shown that, at the linear order in the perturbations, the non-diagonal 
component of the Einstein equations leads to the relation: $\Phi=\Psi$. 
The remaining first order Einstein equations then reduce 
to~\cite{texts,reviews,sasaki}
\br
\!\!\!\!\!\!
\nabla^2 \Phi -3\, {\cal H}\, \l(\Phi^{\prime} 
+{\cal H}\, \Phi\r)
&=& \l(4\pi G\, a^2\r)\, \l[\delta \rho+ {\rho}^{\prime}\, 
\l(B-E^{\prime}\r)\r],\label{eq:foee00}\\
\!\!\!\!\!\!
\pa_{i}\l(\Phi^{\prime}+{\mathcal H}\, \Phi\r)
&=& \l(4\pi G\, a^2\r)\, 
\l[\delta q_{i}+(\rho+p)\; \pa_{i}\l(B-E^{\prime}\r)\r],\\
\!\!\!\!\!\!
\Phi^{\prime\prime}+ 3\, {\mathcal H}\, \Phi^{\prime} 
+\l(2\, {\mathcal H}^{\prime} + {\mathcal H}^2\r)\,\Phi\; 
&=& \l(4\pi G\, a^2\r)\, 
\l[\delta p + p^{\prime}\, \l(B-E^{\prime}\r)\r],
\label{eq:foeeii}
\er
where ${\cal H}=(H\, a)$ is the conformal Hubble parameter, and 
$\delta\rho$, $\delta q_{i}$ and $\delta p$ denote the perturbations 
at the linear order in the energy density, flux, and the pressure of 
the matter field, respectively.
The first and the third of the above first order Einstein equations 
can be combined to lead to the following differential equation for the 
Bardeen potential $\Phi$~\cite{texts,reviews,sasaki}:
\beq
\fl\quad\;
\Phi^{\prime\prime} 
+3\, {\mathcal H}\, \l(1+c_{_{\rm A}}^2\r)\, \Phi^{\prime}
-c_{_{\rm A}}^2\, \nabla^{2}\Phi 
+ \l[2\, {\mathcal H}'+ \l(1+3\, c_{_{\rm A}}^2\r)\, 
{\mathcal H}^2\r]\, \Phi= \l(4\pi G\r)\, a^2\, \delta p_{_{\rm NA}},
\label{eq:emPhi}
\eeq
where we have made use of the standard relation 
\beq
\delta p= c_{_{\rm A}}^2\, \delta\rho +  \delta p_{_{\rm NA}},
\label{eq:deltapgen}
\eeq
with $c_{_{\rm A}}^2\equiv\l(p'/\rho'\r)$ denoting the adiabatic speed
of sound and $\delta p_{_{\rm NA}}$ representing the non-adiabatic 
pressure component.
The quantity $\delta p_{_{\rm NA}}$ is usually related to the entropy 
perturbation ${\cal S}$\footnote{The quantities that appear within 
the square brackets on the right hand sides of the first order Einstein 
equations~(\ref{eq:foee00}) and~(\ref{eq:foeeii}) are the gauge-invariant
versions of $\delta \rho$ and $\delta p$. 
On using these expressions in the relation~(\ref{eq:deltapgen}), it 
is straightforward to show that $\delta p_{_{\rm NA}}$ and, therefore, 
${\cal S}$ are gauge-invariant.} as 
follows (see, for instance, Ref.~\cite{gordon01}):
\beq
\delta p_{_{\rm NA}}=\l(\frac{p'}{\cal H}\r)\, {\cal S}.
\label{eq:deltapna}
\eeq
The curvature perturbation ${\cal R}$ is defined in terms of the 
Bardeen potential $\Phi$ as~\cite{reviews} 
\beq
{\cal R}=\Phi + \l(\frac{2\, \rho}{3\, {\cal H}}\r)\, 
\l(\frac{\Phi'+{\cal H}\, \Phi}{\rho+p}\r).\label{eq:R}
\eeq
On substituting this expression for the curvature perturbation in 
the equation~(\ref{eq:emPhi}) describing the evolution of the 
potential~$\Phi$ and making use of the background equations, we 
obtain that~\cite{gordon01}
\beq
{\cal R}'=-\l(\frac{{\cal H}}{{\cal H}'-{\cal H}^2}\r)\;
\l[\l(\frac{4\pi G\, a^{2}\,p'}{\cal H}\r)\, {\cal S} 
+c_{_{\rm A}}^{2}\; \nabla^{2}\Phi\r].
\label{eq:emR'} 
\eeq

\subsubsection{For the tachyonic case}

To begin with, if we denote the perturbation in the DBI scalar field as 
$\delta T$, then, it is straightforward to show that, the perturbations 
in the energy density, the momentum flux and the pressure of the scalar 
field are given by
\br
\delta \rho 
&=& \l(\frac{V_{T}\, {\delta T}}{\sqrt{1-{\dot T}^{2}}}\r)\,
+\l(\frac{V\, \dot T}{\l(1-{\dot T}^2\r)^{3/2}}\r)\, 
\l({\dot {\delta T}}-A\, {\dot T}\r),\label{eq:deltarho}\\
\delta q_{i} &=& \l(\frac{V\, {\dot T}}{a\, \sqrt{1-{\dot T}^2}}\r) 
\l(\pa_{i}\, \delta T\r),\label{eq:deltaq}\\
\delta p &=& -\l(V_{T}\, \delta T\, \sqrt{1-{\dot T}^2}\r)
+\l(\frac{V\, {\dot T}}{\sqrt{1-{\dot T}^{2}}}\r)\, 
\l({\dot {\delta T}}-A\, {\dot T}\r),\label{eq:deltap}
\er
where $A$ is the quantity that appears in the perturbed Friedmann
line-element~(\ref{eq:frwlesp}).
On substituting these expressions for $\delta\rho$, $\delta q_{i}$ 
and $\delta p$ in the first order Einstein 
equations~(\ref{eq:foee00})--({\ref{eq:foeeii}), we find that the Bardeen 
potential $\Phi$ induced by the tachyonic perturbations satisfies the
following differential equation:
\beq
\fl\quad\;
\Phi^{\prime\prime} 
+3\, {\mathcal H}\, \l(1+c_{_{\rm A}}^2\r)\, \Phi^{\prime}
-c_{_{\rm A}}^2\, \nabla^{2}\Phi 
+ \l[2\, {\mathcal H}'+ \l(1+3\, c_{_{\rm A}}^2\r)\, 
{\mathcal H}^2\r]\, \Phi= \l(c_{_{\rm S}}^2-c_{_{\rm A}}^2\r)\, 
\nabla^{2}\Phi,\label{eq:emPhit}
\eeq
where $c_{_{\rm S}}^{2}=(1-\dot{T}^2)$ is referred to as the effective 
speed of sound (see, for instance, Ref.~\cite{steer04}). 
If we now compare the above equation with Eq.~(\ref{eq:emPhi}) and make 
use of the relation~(\ref{eq:deltapna}), we obtain the corresponding 
entropy perturbation to be~\cite{steer04}
\beq
{\cal S} =\l(\frac{{\cal H}}{4\pi G\, a^2\, p'}\r)\; \l(c_{_{\rm S}}^2
-c_{_{\rm A}}^2\r)\, \nabla^{2}\Phi.\label{eq:St}
\eeq
Therefore, for the tachyonic case, the equation~(\ref{eq:emR'}) describing 
the evolution of the curvature perturbation simplifies to
\beq
{\cal R}'
= -\l(\frac{4\pi G\, a^{2}\,p'}{{\cal H}'-{\cal H}^2}\r)\;
\l(\frac{c_{_{\rm S}}^{2}}{c_{_{\rm S}}^{2}-c_{_{\rm A}}^{2}}\r)\, 
{\cal S}
=  -\l(\frac{{\cal H}\, c_{_{\rm S}}^{2}}{{\cal H}'-{\cal H}^2}\r)\;
\nabla^{2}\Phi.\label{eq:R't}
\eeq
On making use of this relation alongwith the definition~(\ref{eq:R}) and 
the equation~(\ref{eq:emPhit}) describing the evolution of the Bardeen
potential, we find that the Fourier modes of the curvature perturbation 
induced by the DBI scalar field are described by the differential equation 
\beq
{\cal R}_{k}''+2\, \l(\frac{z'}{z}\r)\, {\cal R}_{k}'
+k^{2}\, c_{_{\rm S}}^2\, {\cal R}_k=0,\label{eq:deRk}
\eeq
where the quantity $z$ is given by 
\beq
z = \l(\frac{a}{H}\r)\, \l(\frac{\rho + p}{c_{_{\rm S}}^2}\r)^{1/2}
=\l(\frac{\sqrt{3}\,M_{_{\rm P}}\,a\,\dot{T}}{\sqrt{1 - \dot{T}^2}}\r)
\eeq
and $M_{_{\rm P}}=(8\pi G)^{-1/2}$ denotes the Planck mass.
The scalar power spectrum is then defined as
\beq
{\cal P}_{_{\rm S}}(k)=\l(\frac{k^{3}}{2\, \pi^{2}}\r)\,
\vert{\cal R}_{\rm k}\vert^{2}
\eeq
with the amplitude of the curvature perturbation ${\cal R}_{k}$ evaluated,
in general, at the end of inflation.


\section{Evolution of the curvature perturbations at super-Hubble 
scales}\label{sec:ashsse}

The equation~(\ref{eq:deRk}) that describes the evolution of the 
curvature perturbations is completely equivalent to the equation 
of motion of an oscillator with the time-dependent damping term 
$(z'/z)$.
It is  evident from this correspondence that the curvature 
perturbations can be expected to grow at super-Hubble scales 
(i.e. as $k\to 0$) if there exists a period during which the 
damping term proves to be negative~\cite{leach01a,leach01b}. 
Actually, as we shall see, if $(z'/z)$ turns out to be negative
for some amount of time, then, for a certain range of modes, the 
amplitude of the curvature perturbations is enhanced at 
super-Hubble scales, while the amplitude of another range of modes 
is suppressed, when compared to their values at Hubble exit.
We shall now outline as to how the quantity $(z'/z)$ turns out 
to be negative for the tachyonic case during a period of fast 
roll inflation or when there is a break in inflation.

The slow roll approximation is an expansion in terms of small 
parameters that are defined as derivatives either of the 
potential~$V(T)$ or the Hubble parameter~$H$~\cite{liddle94}.
For our discussion, we shall make use of the horizon flow 
parameters which are defined as the derivatives of the Hubble 
parameter~$H$ with respect to the number of e-foldings~$N$ 
as follows~\cite{schwarz01}:
\beq
\epsilon_0 \equiv \l(\frac{H_*}{H }\r)\quad{\rm and}\quad
\epsilon_{i+1} \equiv \l(\frac{d \ln |\epsilon_i|}{dN}\r)
=\l(\frac{\dot \epsilon_i}{H\, \epsilon_i}\r)\;\; {\rm for}\;\;
i\ge 0, 
\eeq
where $H_*$ is the Hubble parameter evaluated at some given time,
and inflation occurs when $\epsilon_{1}<1$.
It should be pointed out here that these functions form exactly 
the same hierarchy of inflationary flow equations as the Hubble 
slow roll parameters~\cite{liddle94,schwarz01}.

For the case of the DBI scalar field we are considering here, 
the first two horizon flow functions are given by~\cite{steer04}
\beq
\epsilon_{1} = \l(\frac{3\, {\dot T}^2}{2}\r)
\quad{\rm and}\quad
\epsilon_{2} = \l(\frac{2\, \ddot{T}}{H\, \dot{T}}\r).
\label{eq:e1e2}
\eeq
We find that the quantity $(z'/z)$ can be expressed in terms of 
these two horizon flow functions as follows:
\beq
\l(\frac{z'}{z}\r)
=\l(a\, H\r)\, \l(1+\l(\frac{\epsilon_{2}}{2}\r)\, 
\l[\frac{1}{1-\l(2\, \epsilon_{1}/3\r)}\r]\r).\label{eq:z'z}
\eeq
It is apparent from this expression that, in the slow roll limit, 
i.e. when $(\epsilon_{1},\epsilon_{2})\ll 1$, $(z'/z) \simeq (a\,
H) ={\dot a}$ which, in an expanding universe, is a positive 
definite quantity.
Clearly, $(z'/z)$ cannot be negative in the slow roll regime.
However, note that $(z'/z)$ can become negative if the following 
condition is satisfied:
\beq
\epsilon_2 < -2\l[1 - \l(\frac{2\, \epsilon_1}{3}\r)\r].
\label{eq:rc}
\eeq
Such a situation can occur either when there is a break in inflation, 
say, when $\epsilon_{1} = 1$ and $\epsilon_{2}<-(2/3)$ or during 
a period of fast roll inflation, i.e. when $\epsilon_{1}\ll 1$ and 
$\epsilon_{2} < -2$.
In the following subsection, we shall discuss a specific example of 
the second scenario. 

\subsection{A specific example}

To explicitly illustrate the phenomenon of the enhancement or
the suppression of the amplitude of the curvature perturbations 
at super-Hubble scales, we shall work with the potential
\beq
V(T) = V_{0}\, \l( 1+ V_{1}\; T^4 \r),
\eeq
where~$V_{0}$ and~$V_{1}$ are positive constants. 
For this potential, we find that, in the $T$-${\dot T}$ plane, 
there exists only one finite critical point at $(T=0,\, {\dot T}
=0)$ for the equation of motion~(\ref{eq:emT}).
A linear stability analysis about this point immediately suggests 
that the critical point is a stable fixed point.
This implies that, regardless of the initial conditions that the 
tachyon starts rolling from, it will always approach the critical 
point asymptotically. 
Also, we find that, all the trajectories in the phase space rapidly 
approach an attractor trajectory which has the following three regimes 
for the field evolution: ${\dot T} \ll 1$, ${\dot T} \lesssim 1$, and 
${\dot T} \ll 1$ again. 
It is then clear from the expressions~(\ref{eq:e1e2}) for 
$\epsilon_1$ and  $\epsilon_2$  that, as the field starts rolling 
down the potential from a large value, $\epsilon_1$ can grow from 
a small value to some maximum value and then decrease to a small 
value again. 
Also, this behavior will allow $\epsilon_2$ to become negative 
during the period when the field is decelerating---a feature that 
is necessary to achieve the condition~(\ref{eq:rc}).
However, the minimum value attained by $\epsilon_2$ and for how 
long it can remain negative depends on the values of the parameters 
$V_0$ and $V_{1}$ that describe the potential.

We now need to choose the values of the parameters $V_0$ and $V_{1}$ 
of the potential so that the condition~(\ref{eq:rc}) is satisfied. 
A general analysis in terms of the horizon flow functions is not 
possible without explicitly solving the equation of motion for $T$. 
Hence, instead of the horizon flow functions, we shall now make use of the 
potential slow roll parameters to estimate these values~\cite{roberts95}.
For the DBI scalar field we are considering here, the first two 
potential slow roll parameters are defined as~\cite{dbicosmology}
\beq
\varepsilon_{_{\rm V}}
= \l(\frac{M_{_{\rm P}}^2}{2}\r)\, \l(\frac{V_{T}^2}{V^3}\r)
\quad{\rm and}\quad
\delta_{_{\rm V}}
= M_{_{\rm P}}^2\, \l[3\, \l(\frac{V_{T}^2}{V^3}\r) 
-2\, \l(\frac{V_{TT}}{V^2}\r)\r],
\eeq
where $V_{TT}\equiv\l(d^{2}V/dT^{2}\r)$.
For $M_{_{\rm P}}$ set to unity, we find that $\delta_{_{\rm V}}<-2$ 
provided $V_{1}>0.036$\footnote{Actually, the potential slow roll 
parameters and the horizon flow functions are of the same order only 
in the slow roll limit.
Their equivalence will necessarily break down in the fast roll regime 
we are interested in.
However, we find that the value for $V_{1}$ we shall work with (which is
greater than $0.036$) indeed leads to the required behavior.}.

We have solved the background equations and the equation describing the 
curvature perturbation numerically. 
The results that we present here are for the following values of the
potential parameters: $V_{0}=0.5$ and $V_{1}= 0.18$.
We have chosen the standard initial conditions at sub-Hubble scales 
corresponding to the Bunch-Davies vacuum for the curvature perturbations. 
The initial conditions we have imposed are easily expressed in terms of 
the Mukhanov-Sasaki variable $v_{k}$ that is related to the curvature 
perturbation through the relation: $v_{k}=({\cal R}_{k}\, z)$.
The initial conditions we have imposed are as follows (see, for instance, 
the second textbook in Ref.~\cite{texts}):
\beq
v_{k}=\l(\frac{1}{2\, \omega_{k}}\r)^{1/2}
\qquad{\rm and}\qquad
v_{k}'=-i\, \l(\frac{\omega_{k}}{2}\r)^{1/2},
\eeq
where 
\beq
\omega_{k}^{2}=\l[\l(k\, c_{_{\rm S}}\r)^{2} -\l(z''/z\r)\r],
\eeq
and we have imposed these conditions at a given time when all the
modes of interest are well inside the Hubble radius.

In Figure~\ref{fig:z1z}, we have plotted the quantity $(z'/z)$ as a 
function of the number of e-foldings~$N$. 
It is clear from the figure that $(z'/z)$ is negative during $58<N<61$.
\begin{figure}
\begin{center}
\vskip 25pt
\resizebox{270pt}{180pt}{\includegraphics{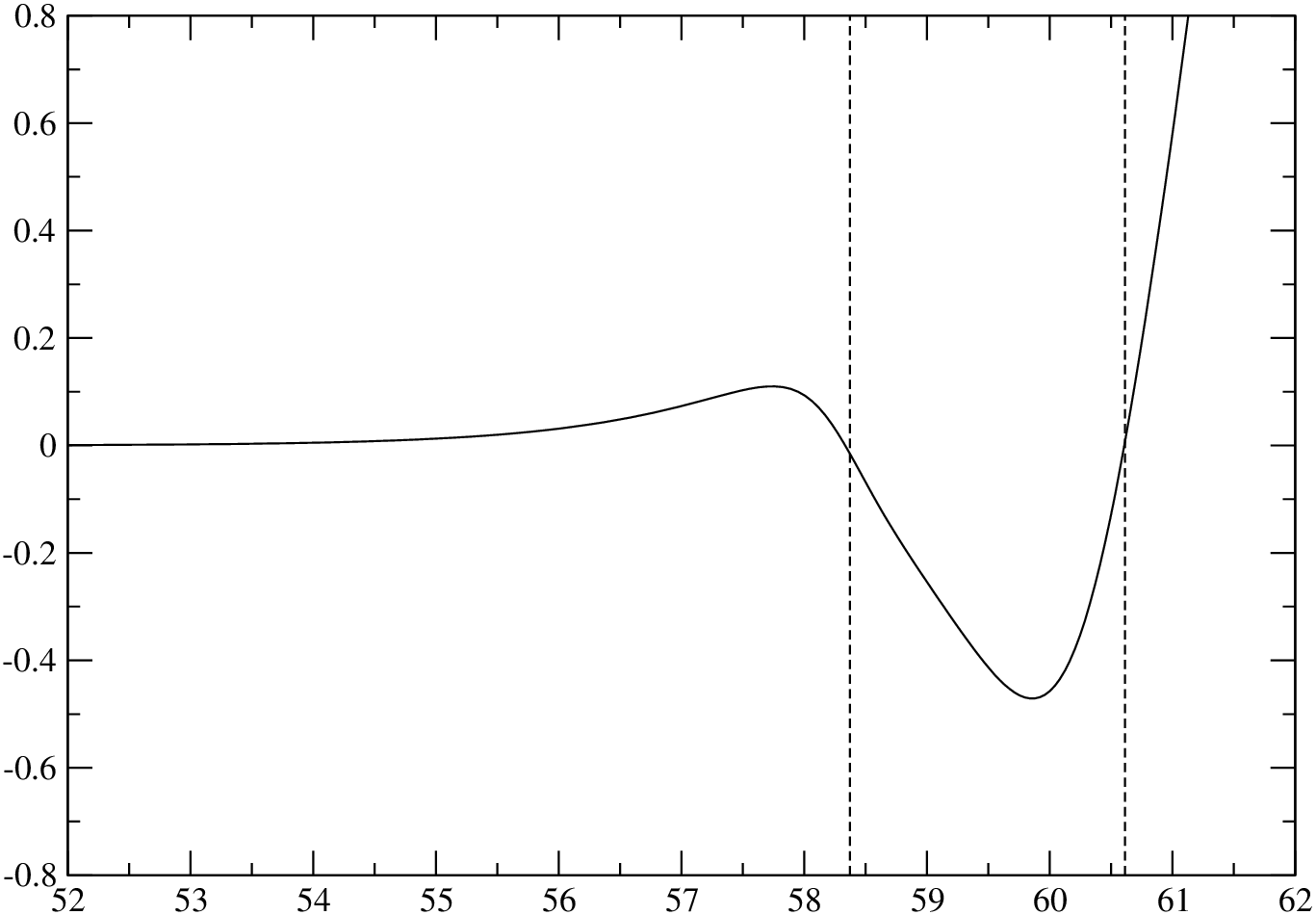}}
\vskip -105 true pt \hskip -320 true pt $(z'/z)$
\vskip 95 true pt \hskip 15 true pt $N$
\vskip 2 true pt
\caption{The evolution of the quantity $(z'/z)$ has been plotted as a 
function of the number of e-folds $N$. 
The vertical lines indicate the regime where $(z'/z)$ is negative.
Note that it remains negative for a little less than three e-folds 
between~$N$ of~$58$ and~$61$.}\label{fig:z1z}
\end{center}
\end{figure}
In Figure~\ref{fig:Rk}, we have plotted the amplitude of the 
curvature perturbation ${\cal R}_{k}$ as a function of $N$ for two 
modes which are at super-Hubble scales when the slow roll to fast 
roll transition takes place.
\begin{figure}
\begin{center}
\vskip 25pt
\resizebox{270pt}{180pt}{\includegraphics{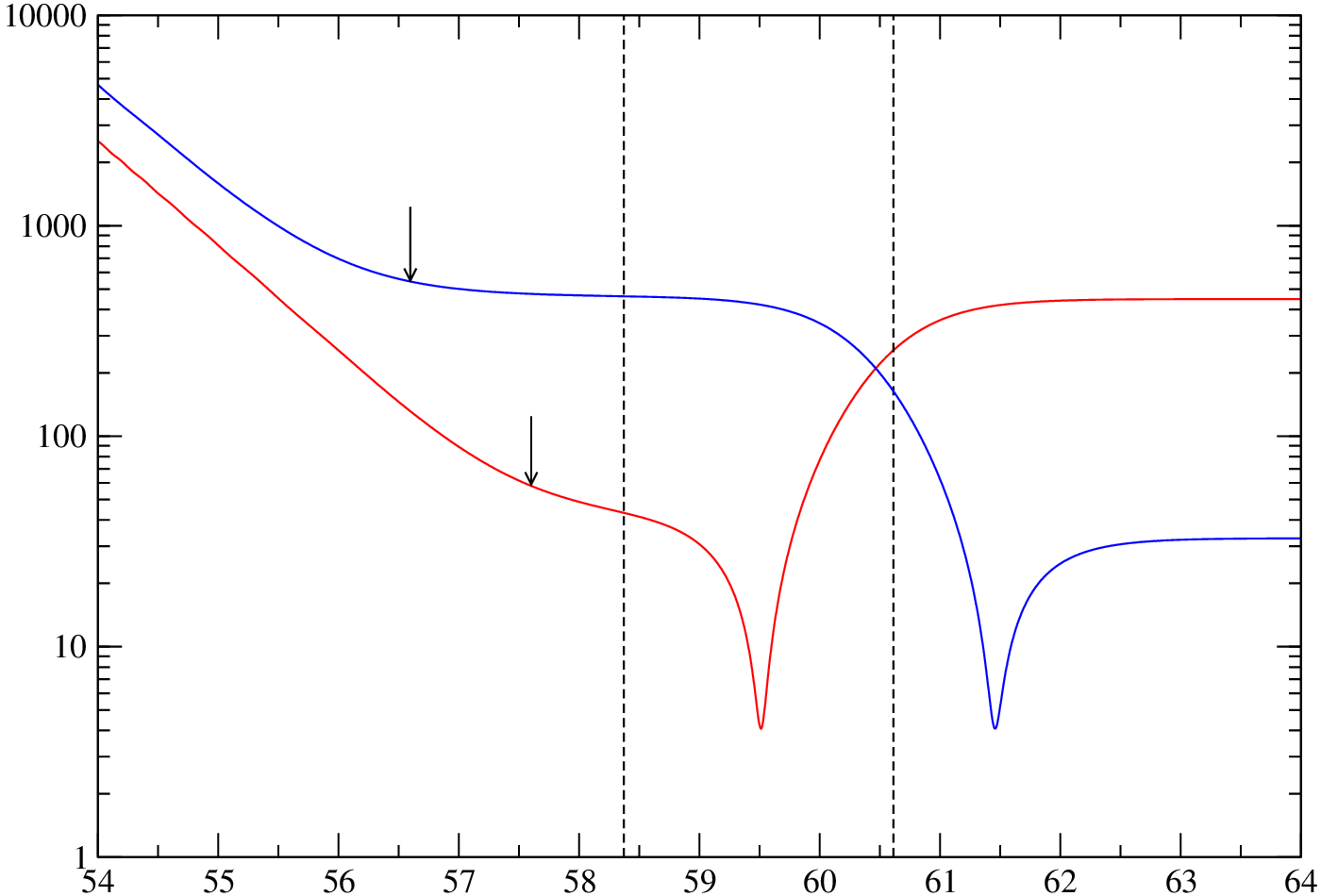}}
\vskip -105 true pt \hskip -300 true pt {$\vert{\cal R}_{k}\vert$}
\vskip 95 true pt \hskip 15 true pt $N$
\vskip 2 true pt
\caption{The evolution of the amplitude of the curvature perturbation 
${\cal R}_{k}$ is plotted as a function of the number of e-folds $N$ 
for the modes with wave numbers $k=0.03$ (in blue) and  $k=0.1$ (in red).
As in the previous figure, the vertical lines delineate the regime 
where $(z'/z)$ is negative.
The arrows indicate the time at which the modes leave the Hubble 
radius, i.e. when $(k\, c_{_{\rm S}})=(a\, H)$.
Note that the amplitude of the curvature perturbation for the $k=0.1$ 
mode is enhanced at super-Hubble scales (by a factor of about $10$), 
while the amplitude of the $k=0.03$ mode is suppressed (by a factor of
about $20$), when compared to their values at Hubble exit.
We should mention that these two particular modes have been chosen 
for the reason that, in the specific example that we are considering, 
they exhibit sufficient extent of amplification or suppression at 
super-Hubble scales.}
\label{fig:Rk}
\end{center}
\end{figure}
It is evident from the figure that, while the amplitude of
${\cal R}_{k}$ corresponding to the mode with wave number 
$k=0.1$ is enhanced at super-Hubble scales, the amplitude
of the mode with wave number $k=0.03$ is suppressed at late
times, when compared to their values at Hubble exit\footnote{In 
Refs.~\cite{leach01a,leach01b}, the authors emphasize the point 
that, a period of deviation from slow-roll inflation can enhance
the amplitude of the curvature perturbations at super-Hubble 
scales, but seem to overlook the fact that it can also lead 
to their suppression. 
Actually, they also encounter the suppression of the amplitude
in the examples they consider.
In Figures~1 and~2 of Ref.~\cite{leach01b}, they plot the power 
spectrum evaluated at the end of inflation as well as at a time
soon after the modes leave the Hubble radius.
It is clear from these two figures that, while a certain range of 
modes are amplified at super-Hubble scales, another range of modes 
are suppressed.}.
We should mention that, to highlight these behaviors, we have chosen 
modes that exhibit sufficient amplification or suppression for the 
specific model and parameters we are considering here.
Finally, we should also stress the point that, had there been no 
transition to the fast roll regime, the amplitude of the curvature 
perturbations would have frozen at their value at Hubble exit. 


\subsection{Entropy perturbations as the source of evolution
at super-Hubble scales}

In this subsection, we shall outline as to how the entropy 
perturbation~${\cal S}$ grows during the fast roll regime, and how such 
a growth in turn acts as the source for the change in the amplitude of 
the curvature perturbation at super-Hubble scales.

On using the expression~(\ref{eq:St}) for the entropy perturbation 
${\cal S}$, we find that the differential equations~(\ref{eq:emR'}) 
and (\ref{eq:deRk}) describing the evolution of~${\cal R}$ can be
written in Fourier space as the following two first order 
equations~\cite{leach01a}:
\br
\l(\frac{{\cal R}_{k}'}{aH}\r) 
&=&  {\cal A}\, {\cal S}_{k},\label{eq:R'}\\
\l(\frac{{\cal S}_{k}'}{aH}\r) 
&=& {\cal B}\; {\cal S}_{k} 
- {\cal C}\, \l(\frac{k^2}{a^2H^2}\r)\, {\cal R}_{k}.
\label{eq:S'}
\er
The quantities ${\cal A}$, ${\cal B}$ and ${\cal C}$ appearing 
in the above two equations can be expressed in terms of the 
first three horizon flow parameters as
\br
\fl\quad
{\cal A} = \l(3-2\,\epsilon_1\r)\,
\l(\frac{3 - 6\,\epsilon_1 + \epsilon_2}{6 
-12\, \epsilon_1  +\epsilon_2}\r),\label{eq:A}\\
\fl\quad
{\cal B} = \l(\frac{1}{(3-2\epsilon_1)\, \l(6
-12\, \epsilon_1+ \epsilon_2\r)\, 
\l(3 -6\, \epsilon_1+ \epsilon_2\r)}\r)\nn\\
\!\!\!\!\!\!\!\!\!\!\!\!\!\!\!\!
\times\,\biggl[\l(2\,\epsilon_1 \epsilon_2\, \l[3 
- 72\,\epsilon_1^{2}+ \epsilon_2\, \l(12+\epsilon_2\r)
- 6\, \epsilon_1\, \l(12+2\, \epsilon_2 + \epsilon_3\r)\r] 
- 9\, \epsilon_2 \epsilon_3\r)\nn\\
\qquad\qquad\;\;- \l(6-12\, \epsilon_1+\epsilon_2\r)\, 
\l(3  -6\,\epsilon_1+ \epsilon_2\r)\, 
\l(9 - 9\,\epsilon_1  + 3\, \epsilon_2 + 2\, \epsilon_1^2\r)\biggr],
\label{eq:B}\\
\fl\quad
{\cal C} = \l(\frac{1}{3}\r)\, \l(\frac{6-12\, \epsilon_{1}
+\epsilon_{2}}{3- 6\,\epsilon_{1}+\epsilon_{2}}\r),\label{eq:C}
\er
with the third parameter $\epsilon_{3}$ given by
\beq
\epsilon_3 
=\l(\frac{1}{H}\r)\, \l[\l(\frac{\tdot{T}}{\ddot T}\r)
-\l(\frac{\ddot T}{\dot T}\r)-\l(\frac{\dot H}{H}\r)\r].
\label{eq:hfp3}
\eeq
We shall now make use of the coupled first order differential
equations~(\ref{eq:R'}) and~(\ref{eq:S'}) to understand the 
evolution of the entropy perturbations at super-Hubble scales 
and its effect on the curvature perturbations during the slow 
roll and the fast roll regimes.

Let us first discuss the behavior in the slow roll regime.
During slow roll, we can ignore the horizon flow parameters when 
compared to the numerical constants in the above expressions for 
${\cal A}$, ${\cal B}$ and ${\cal C}$.
Since we are interested in the evolution at super-Hubble scales, 
one would be tempted to ignore the term involving ${\cal R}_{k}$
in Eq.~(\ref{eq:S'}).
If we can indeed do so, we can immediately conclude that ${\cal S}_{k} 
\propto e^{-3N}$ during an epoch of slow roll inflation.
However, the self-consistent numerical solutions we have obtained 
to equations~(\ref{eq:R'}) and~(\ref{eq:S'}) indicate otherwise. 
In Figure~\ref{fig:Sk}, using the numerical solutions to the 
curvature perturbations ${\cal R}_{k}$ we had discussed in the 
last section and the relation~(\ref{eq:R'}), we have plotted the 
evolution of the entropy perturbation ${\cal S}_{k}$ for the two 
modes $k=0.03$ and $k=0.1$ as a function of the number of e-folds.
It is clear from the figure that while there is indeed an intermediate
period in the slow roll phase when ${\cal S}_{k} \propto e^{-3N}$, the 
late time behavior actually has the form ${\cal S}_{k} \propto e^{-2N}$.
In fact, we find that all the modes that leave the Hubble radius
before the transition to fast roll exhibit such a behavior.
We should point out here that similar conclusions have been arrived 
at earlier for the case of perturbations induced by the canonical 
scalar field~\cite{leach01a}.

During a period of fast roll inflation, in contrast to the slow roll 
case, we cannot ignore the horizon flow parameters in the quantities 
${\cal A}$, ${\cal B}$ and ${\cal C}$.
If we now choose to neglect the term involving ${\cal R}_{k}$ in 
Eq.~(\ref{eq:S'}), we find that the equation reduces to
\beq
\l(\frac{{\cal S}_{k}'}{aH}\r) \simeq 
\l({\cal B}\, {\cal S}_{k}\r). 
\eeq
Moreover, we find that, on using the background equations, we 
can rewrite the expression~(\ref{eq:hfp3}) for $\epsilon_{3}$ 
as follows:
\br
\!\!\!\!\!\!\!\!\!\!\!\!\!\!\!\!\!\!\!\!\!\!\!\!
\epsilon_3 
&=& -\l(\frac{3}{2}\r) + \l(\frac{7\, \epsilon_1}{2}\r) 
- \l(\frac{\epsilon_2}{4}\r) 
-\l(\frac{\epsilon_1\, \epsilon_2}{2(3-2\epsilon_1)}\r)\nn\\
\!\!\!\!\!\!\!\!\!\!\!\!\!\!\!\!\!\!\!\!\!\!\!\!
& &\qquad\qquad\qquad\qquad\qquad\;\;\,
+ \l(\frac{3\, \epsilon_1}{\epsilon_2}\r)  - \l(\frac{V_{TT}}{VH^2}\r)\,
\l(\frac{1}{\epsilon_2}\r)\, \l[1 - \l(\frac{2\, \epsilon_1}{3}\r)\r].
\er
If we now assume that $\epsilon_1\ll 1$ and make the additional 
assumption that $\l(V_{TT}/V\, H^2\r) \ll 1$\footnote{We should 
point out here that, for the case of the standard scalar field, 
the quantity $(V_{TT}/H^2)$ is actually the second potential slow 
roll parameter.
However, for the DBI field, the equivalent quantity turns out to be 
$(V_{TT}/V\, H^2)$, and the extra factor of $V$ in the denominator is 
due to the form of the DBI action.}, we 
find that $\epsilon_3$ simplifies to~\cite{leach01a}
\beq
\epsilon_3 \simeq -\l(\frac{3}{2}\r) - \l(\frac{\epsilon_2}{4}\r).
\eeq
On substituting this value in the expression~(\ref{eq:B}) for 
${\cal B}$ and assuming that $\epsilon_2 \simeq -5$ (which is 
roughly the largest value of $\epsilon_2$ in the fast roll 
regime for our choice of parameters), we obtain that 
\beq
{\cal S}_{k} \propto e^{4N}.  
\eeq
As we have illustrated in Figure~\ref{fig:Sk}, this rough estimate 
is corroborated by the numerical result we obtain. 
\begin{figure}[!htb]
\begin{center}
\vskip 25pt
\resizebox{270pt}{180pt}{\includegraphics{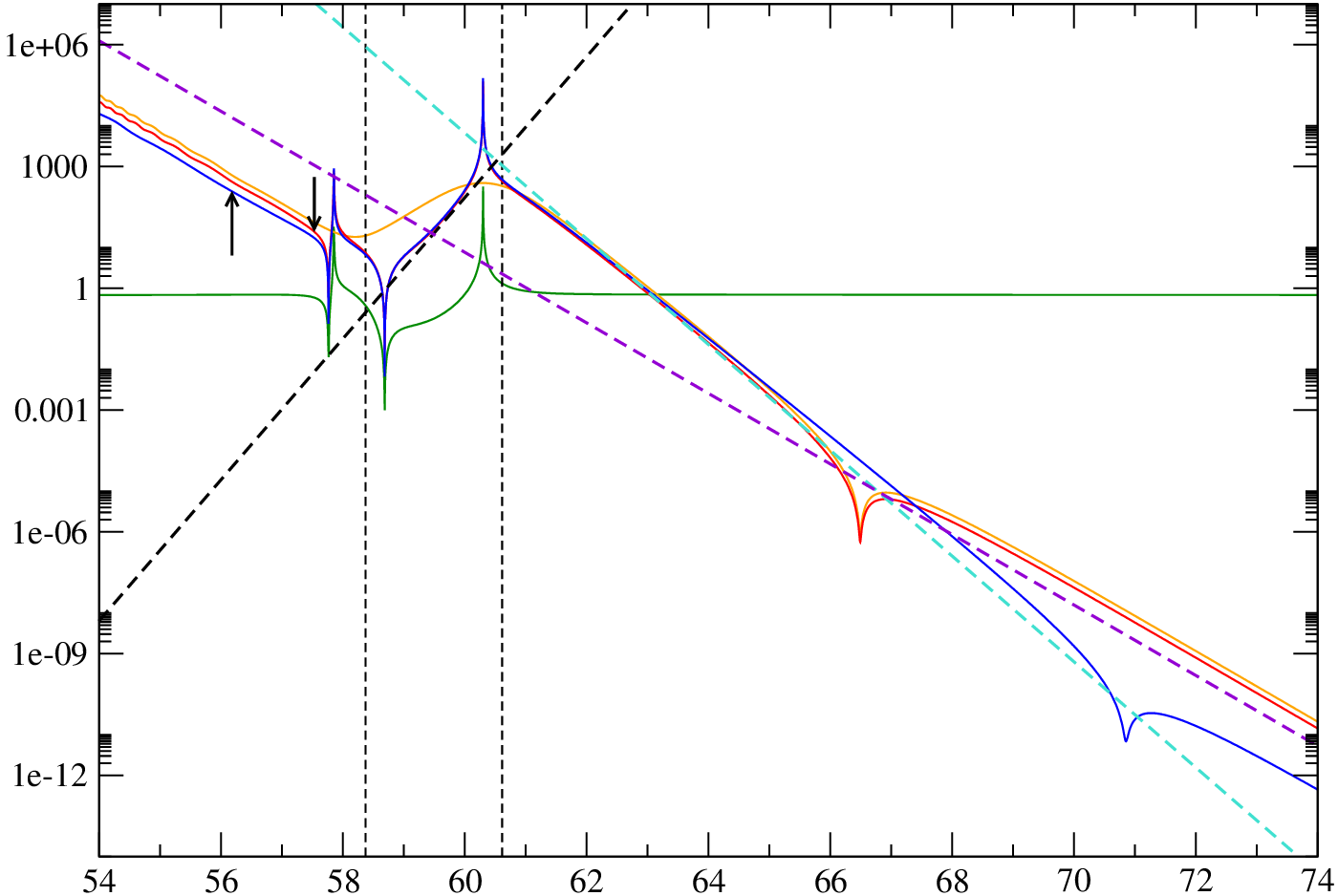}}
\vskip -105 true pt \hskip -300 true pt {$\vert{\cal S}_{k}\vert$}
\vskip 95 true pt \hskip 15 true pt $N$
\vskip 2 true pt
\caption{The evolution of the amplitude of the entropy perturbation 
${\cal S}_{k}$ is plotted as a function of the number of e-folds $N$ 
for the two modes $k=0.03$ (in blue) and  $k=0.1$ (in red) we had
considered in the previous figure. 
The dashed lines in black, turquoise and violet indicate the 
$e^{(4N)}$, $e^{-(3N)}$ and $e^{-(2N)}$ behavior, respectively.  
The vertical lines again delineate the fast roll regime and, as 
before, the arrows indicate the time at which the modes leave the 
Hubble radius.
We have also plotted the quantities $(1/{\cal A})$ (in green)
and $(\vert{\cal R}_{k}'\vert/a\, H)$ (in orange) for the mode  
$k=0.1$.
The former is discontinuous while the latter is continuous, 
and it is the former quantity that leads to the discontinuities 
in the evolution of ${\cal S}_{k}$.
The conclusions we have discussed in the text---the  $e^{(4N)}$ growth
of the entropy perturbation during fast roll, the intermediate $e^{-(3N)}$ 
slow roll behavior and the late time $e^{-(2N)}$ slow roll decay---are 
evident from the figure.
It is also important to note that the entropy perturbations associated
with both the modes evolve in a similar fashion during the fast roll
phase.}
\label{fig:Sk}
\end{center}
\end{figure}
It is then evident from equation~(\ref{eq:R'}) that it is such a
rapid growth of the entropy perturbation during the fast roll regime 
(instead of a slow roll decay) that is responsible for the change 
in the amplitude of the curvature perturbation at super-Hubble 
scales.
However, we should add that a rapid growth of entropy perturbations 
is exhibited only by modes that leave the Hubble radius just before 
the transition to the fast roll phase.
We find that the earlier the modes leave the Hubble radius, less
rapidly do their entropy perturbations grow during the fast roll 
phase.
Evidently, the longer a mode has been outside the Hubble radius 
before the fast roll transition, the more suppressed the entropy 
perturbation is, and lesser is its growth during the fast roll 
phase. 
Therefore, its ability to affect the curvature perturbation gets 
suppressed correspondingly.

The following clarifying remarks are in order at this stage of our 
discussion.
To begin with, we should reiterate the point we had made earlier, 
viz. that, in the absence of a transition to the fast roll regime, 
the amplitude of the curvature perturbations would have frozen at 
their value at Hubble exit. 
Also, since, at sub-Hubble scales, the modes do not feel the effect 
of the background quantities, the transition has virtually no effect 
on those modes that are well within the Hubble radius during the period 
of fast roll inflation. 
In fact, the fast roll regime has the maximum effect on the modes that 
leave the Hubble radius just before the transition.
(As we mentioned, it is for this reason that, in Figures~\ref{fig:Rk}
and~\ref{fig:Sk}, we have chosen modes that leave the Hubble scale 
just before the transition to the fast roll regime takes place and 
which exhibit sufficient extent of amplification or suppression.)
While the amplitude of a certain range of modes is indeed enhanced at
super-Hubble scales as has been noticed earlier in the case of the 
canonical scalar field~\cite{leach01a}, we find that, actually, there 
also exists a range of modes whose amplitude is {\it suppressed}\/ at 
super-Hubble scales, when compared to their value at Hubble exit.
This is evident from Figure~\ref{fig:ps} wherein we have plotted 
the power spectrum evaluated soon after Hubble exit as well as the 
spectrum computed at the end of inflation.
Moreover, the extent of the change in the amplitude of the curvature 
perturbations proves to be smaller and smaller for modes that leave 
the Hubble radius earlier and earlier before the transition.
Clearly, this is due to the combined effect of the slower growth 
during the fast roll phase and the exponential suppression of the 
entropy perturbations far outside the Hubble radius.


\section{The long wavelength approximation}\label{sec:cep}

In the previous section, working in the fluid picture, we had 
discussed as to how the growth of entropy perturbations acts as 
the source for the evolution of the curvature perturbations at 
super-Hubble scales during a period of fast roll inflation.
In this section, extending the earlier result for the canonical 
scalar field~\cite{leach01b}, we evaluate the power spectrum in
the long wavelength approximation and show that it matches the
exact spectrum obtained numerically, quite well.
Since the analysis for the DBI field is essentially similar to that 
of the canonical scalar field, rather than repeat the discussion, 
we shall simply point out the difference between the two cases
and present the final results.

As far as the evolution of the curvature perturbation goes, the only 
difference between the canonical and the DBI scalar fields is the 
factor of $c_{_{\rm S}}^{2}$ that appears as the coefficient of $k^2$
in Eq.~(\ref{eq:deRk}).
The effective speed of sound $c_{_{\rm S}}$ turns out to be unity for
the standard scalar field.
In the earlier analysis for the canonical scalar field~\cite{leach01b},
the amplitude of curvature perturbation at the end of inflation was 
related to its value soon after Hubble exit in terms of quantities 
involving the function $z$ at the ${\cal O}(k^2)$ in the long wavelength
approximation.
Therefore, the corresponding result for the DBI scalar field
essentially involves suitably replacing the quantity $k^2$ with 
$(k^{2}\, c_{_{\rm S}}^{2})$.
We find that, at the ${\cal O}(k^2)$, the amplitude of the curvature 
perturbation at the end of inflation, say, at $\eta_{\ast}$, can be 
related to its amplitude soon after Hubble exit, say, at $\eta_{k}$, 
by the following relation:
\beq
{\cal R}_{k}(\eta_{\ast})=\l[\alpha_{k}\, {\cal R}_{k}(\eta_{k})\r],
\label{eq:Rklwa}
\eeq
where $\alpha_{k}$ is given by
\beq
\alpha_{k}=\l[1+D_{k}(\eta_{k})-F_{k}(\eta_{k})\r].
\eeq
The quantities $D_{k}$ and $F_{k}$ in the above expression for 
$\alpha_{k}$ are described by the following integrals involving
the functions $z(\eta)$ and $c_{_{\rm S}}(\eta)$:
\beq
D_{k}(\eta)
\simeq {\cal H}_k\, \int\limits_{\eta}^{\eta_*}\!
d\eta_{1}\, \l(\frac{z^2(\eta_k)}{z^2(\eta_{1})}\r)
\eeq
and
\beq
F_{k}(\eta) 
\simeq k^2\, 
\int\limits_{\eta}^{\eta_{*}}\! \frac{d\eta_{1}}{z^2(\eta_{1})}\;
\int\limits^{\eta_{1}}_{\eta_{k}}\! d\eta_{2}\, \l[c_{_{\rm S}}^2(\eta_{2})\; 
z^2(\eta_{2})\r]
\eeq
with ${\cal H}_{k}$ denoting the conformal Hubble parameter evaluated
at $\eta_{k}$.

As we had pointed out before, during slow roll inflation, $(z'/z)\simeq 
(a\, H)={\dot a}$.
In other words, in the slow roll regime, $z$ is a monotonically increasing
quantity.
Therefore, in such a regime, the quantity $D_{k}$ remains small and the 
contribution due to $F_{k}$---since it is proportional to $k^2$---can be 
ignored at extreme super-Hubble scales.
However, the contributions due to these two terms cannot be neglected 
if these quantities turn out to be larger than unity.
Recall that $c_{_{\rm S}}^{2}=(1-{\dot T}^{2})$ and, hence, 
$0\le c_{_{\rm S}}^{2}\le 1$.
It is then clear from the above integrals for $D_{k}$ and $F_{k}$  
that they can be large if, for a given mode, there exists an epoch 
during which $z(\eta)$ at super-Hubble scales is much smaller than 
the corresponding value when the mode left the Hubble radius.
We find that indeed such a situation arises during the fast roll 
regime in the specific example we had discussed in the last section.
For the potential and the parameters we were working with, 
we find that $0.65< c_{_{\rm S}}^{2}< 1$.
And, in Figure~\ref{fig:z2}, we have plotted the quantity $z^{2}$ as 
a function of the number of e-foldings.
It is the dip in the quantity $z^{2}$ during the fast roll regime that 
turns out to be responsible for the change in the amplitude of the 
curvature perturbation at super-Hubble scales.
\begin{figure}[!htb]
\begin{center}
\vskip25pt
\resizebox{270pt}{180pt}{\includegraphics{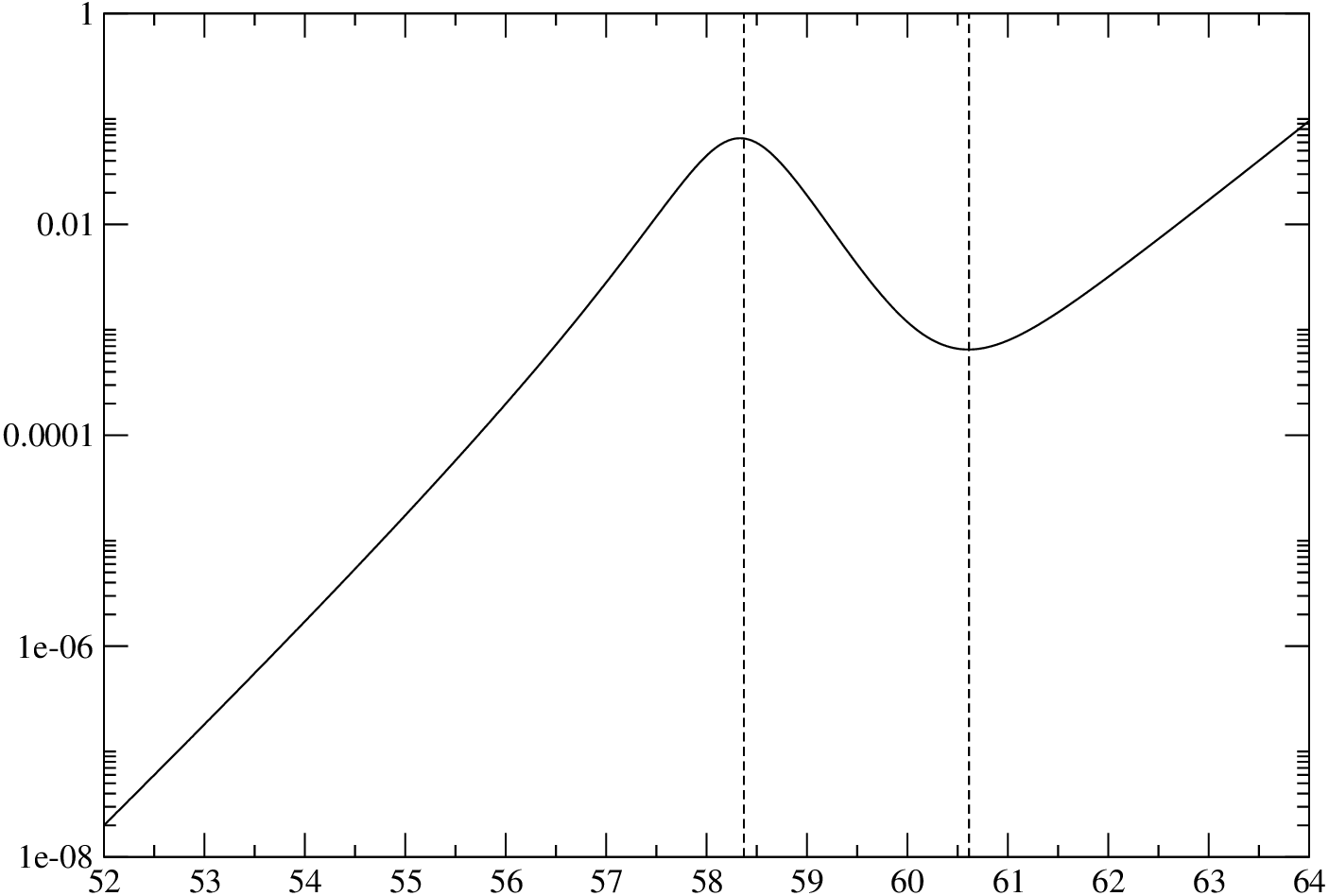}}
\vskip -105 true pt \hskip -320 true pt $z^{2}$
\vskip 95 true pt \hskip 15 true pt $N$
\vskip 2 true pt
\caption{The evolution of the quantity $z^{2}$ is plotted as a function 
of the number of e-folds $N$.
Note the dip during the fast roll regime that is outlined by the vertical
lines.
It is this dip that is responsible for $(z'/z)$ being negative, which in 
turn leads to the change in the amplitude of the curvature perturbations 
at super-Hubble scales.}\label{fig:z2}
\end{center}
\end{figure}

In Figure~\ref{fig:ps}, using the numerical integration of the 
modes we had discussed in the last section, we have plotted the 
power spectrum evaluated at the end of inflation as well as the 
spectrum that has been obtained in the long wavelength 
approximation using the relation~(\ref{eq:Rklwa}).
We have also plotted the power spectrum evaluated soon after the 
modes leave the Hubble radius.
\begin{figure}[!htb]
\begin{center}
\vskip25pt
\resizebox{270pt}{180pt}{\includegraphics{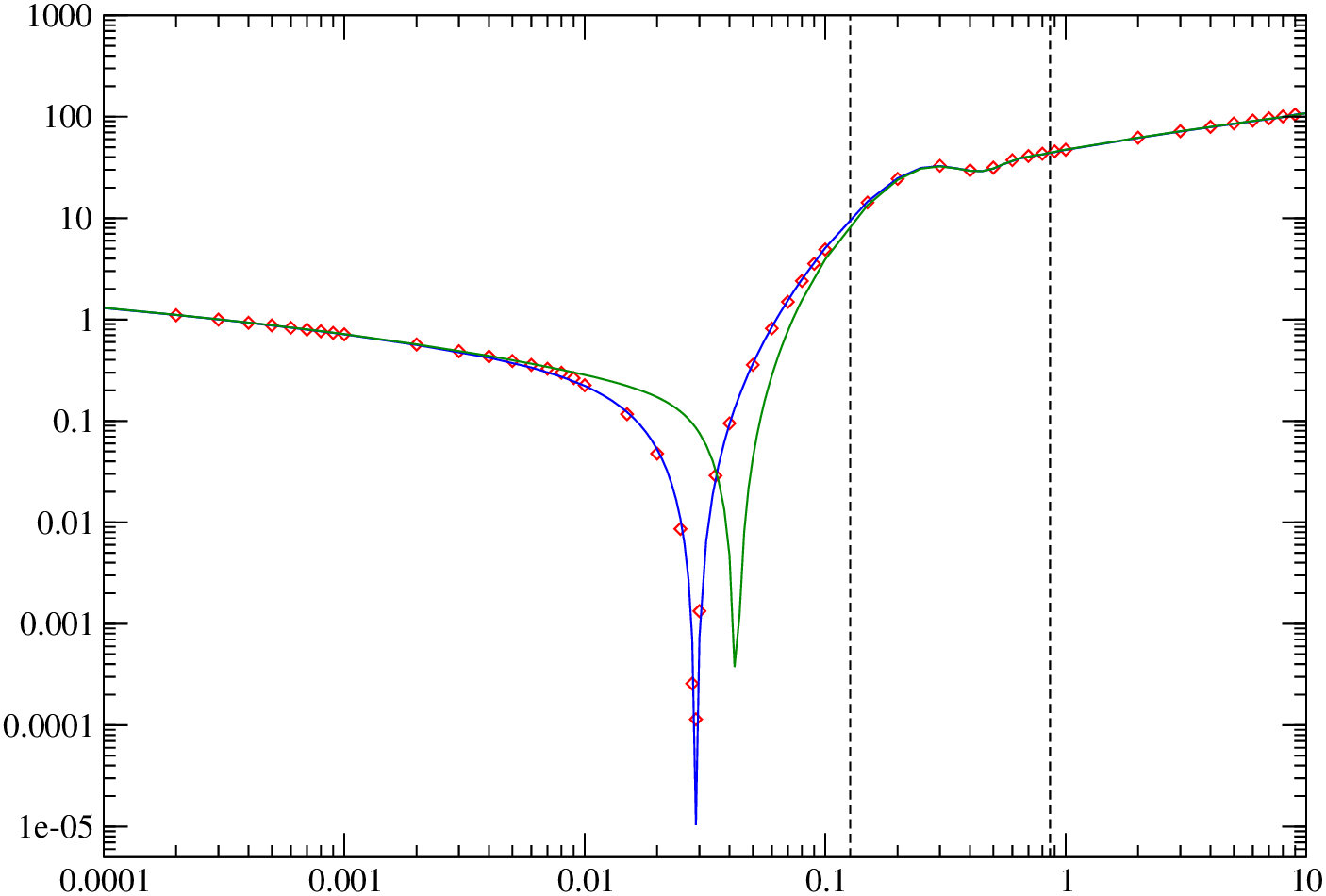}}
\vskip -105 true pt \hskip -320 true pt ${\cal P}_{_{\rm S}}(k)$
\vskip 95 true pt \hskip 15 true pt $k$
\vskip 2 true pt
\caption{Plots of the scalar power spectrum that has been evaluated 
at the end of inflation through the numerical integration of the modes 
(in blue) as well as the spectrum that has been obtained using the 
long wavelength approximation (in red).
The spectrum evaluated soon after the modes leave the Hubble radius 
(when $(k\, c_{_{\rm S}}/a\, H)=0.1$) has also been plotted (in green).
The modes within the vertical lines leave the Hubble radius 
during the fast roll regime.
It is clear from the figure that the power spectrum evaluated in the
long wavelength approximation agrees quite well with the actual
spectrum.
However, there is a substantial difference between the power spectrum 
evaluated near Hubble exit and the spectrum evaluated at the end of 
inflation.
The difference is, in particular, large for those modes whose Hubble 
exit occurs just before the period of fast roll inflation.
Note that, while the amplitude of the modes at super-Hubble scales in 
the wave number range $0.01\lesssim k \lesssim 0.04$ is suppressed 
when compared to its value near Hubble exit, the amplitude of the modes 
in the range $0.04\lesssim k \lesssim 0.1$ is enhanced.
Essentially, the valley in the power spectrum computed near Hubble 
exit has shifted towards smaller $k$ when the spectrum is evaluated 
in the super-Hubble limit.}
\label{fig:ps}
\end{center}
\end{figure}
The figure clearly illustrates the following three points.
Firstly, the spectrum evaluated at the ${\cal O}(k^2)$ in the long 
wavelength limit proves to be quite a good fit of the actual spectrum.
Secondly, there is a considerable difference between the power spectrum 
that has been evaluated at the end of inflation and the spectrum that 
has been evaluated soon after Hubble exit.
The difference is, in particular, large for the modes that leave the 
Hubble radius just before the fast roll regime.
Thirdly, while the amplitude of the modes at super-Hubble scales in 
the wave number range $0.01 \lesssim k \lesssim 0.04$ is suppressed 
when compared to its value soon after Hubble exit, the amplitude of 
the modes in the range $0.04 \lesssim k \lesssim 0.1$ is enhanced.


\section{Summary}\label{sec:summary}

In this work, with the help of a specific example, we have shown 
that the amplitude of the curvature perturbations induced by a DBI 
scalar field can be {\it enhanced or suppressed}\/ at super-Hubble 
scales if there exists a period of deviation from slow roll inflation.
Working in the fluid picture, we have illustrated that, as in the
case of the canonical scalar field, the change in the amplitude of 
the curvature perturbations arises due to the growth of the entropy 
perturbations in the fast roll regime.
Moreover, following the results obtained earlier for the standard 
scalar field, we have shown that the power spectrum evaluated in 
the long wavelength approximation matches the exact spectrum 
obtained numerically very well. 

Transitions from slow roll to fast roll inflation lead to deviations 
from a nearly scale-invariant power spectrum. 
The possibility that specific deviations from the standard scale 
independent power spectrum may fit the cosmic microwave background 
observations better have been explored recently in the 
literature~\cite{contaldi,covi}. 
If we can systematically understand the effects of the quantity 
$z$ on the curvature perturbation ${\cal R}_{k}$, we will be able to 
fine tune the transitions such as the ones we have discussed in this 
work to obtain the desired features in the power spectrum (in this 
context, see Ref.~\cite{covi}). 
For instance, the sharp drop in the power spectrum in our model in the
wavelength range $0.03 \lesssim k \lesssim 0.1$ may provide a better 
fit to the lower power observed in the quadrupole moment of the cosmic 
microwave background~\cite{tarun}.
We are currently investigating these issues.


\ack{The authors would wish to thank Kim Juhan, T.~Padmanabhan, 
S.~Shankaranarayanan and K.~P.~Yogendran for discussions.
We would also like to acknowledge the use of the cluster computing 
facility at the Harish-Chandra Research Institute (HRI), Allahabad, 
India.
PC would also like to thank HRI for hospitality where part of this 
work was carried out.}


\end{document}